\documentstyle [eqsecnum,preprint,aps,epsf,floats,fixes]{revtex}
\tightenlines
\begin {document}

\def\D{{\bf D}}
\def\barD{{\bf\bar D}}
\def\E{{\bf E}}

\def\B{{\bf B}}
\def\barB{{\bf\bar B}}
\def\A{{\bf A}}
\def\barA{{\bf\bar A}}
\def\J{{\bf J}}
\def\x{{\bf x}}
\def\y{{\bf y}}
\def\z{{\bf z}}
\def\p{{\bf p}}
\def\k{{\bf k}}
\def\q{{\bf q}}
\def\v{{\bf v}}

\def\g{{\rm g}}
\def\grad{\mbox{\boldmath$\nabla$}}
\def\noise{\zeta}
\def\bnoise{\mbox{\boldmath$\noise$}}
\def\barbnoise{\mbox{\boldmath$\bar\noise$}}
\def\noiseA{\xi}
\def\CA{C_{\rm A}}
\def\nf{n_{\rm f}}
\def\pl{{\rm pl}}
\def\mD{m_{\rm D}}
\def\md{m_{\rm D}}

\def\I{{\cal I}}

\def\eff{{\rm eff}}
\def\tr{{\rm tr}}
\def\Tr{{\rm Tr}}

\def\Prob{{\cal P}}
\def\PL{P_{\rm L}}
\def\PLpert{P_{\rm L}^{\rm pert}}
\def\PT{P_{\rm T}}

\def\V{{\cal V}}

\def\eps{\epsilon}

\def\deltaS{\delta^{(2)}}
\def\r{{\bf r}}
\def\barr{{\bf\bar r}}
\def\Pr{P_{\rm r}}
\def\eff{{\rm eff}}
\def\btheta{\mbox{\boldmath$\theta$}}

\def\half{{\textstyle{1\over2}}}
\def\third{{\textstyle{1\over3}}}
\def\fourth{{\textstyle{1\over4}}}

\preprint {UW/PT 98-11}

\title  {Longitudinal subtleties in diffusive Langevin equations for
         non-Abelian plasmas}

\author {Peter Arnold}

\address
    {%
    Department of Physics,
    University of Virginia,
    Charlottesville, VA 22901
    }%
\author{Dam T. Son}
\address
    {%
    Center for Theoretical Physics,
    Department of Physics, \\
    Massachusetts Institute of Technology,
    Cambridge, MA 02139
    }%
\author{Laurence G. Yaffe}
\address
    {%
    Department of Physics,
    University of Washington,
    Seattle, Washington 98195
    }%
\date {January 1999}

\maketitle
\vskip -20pt

\begin {abstract}%
{%
B\"odeker has recently argued that non-perturbative processes in very high
temperature non-Abelian plasmas (such as electroweak baryon number violation
in the very hot early Universe) can be quantitatively described,
to leading logarithmic accuracy, by a simple diffusive effective theory.
B\"odeker's effective theory is intended to describe the long-distance
transverse electric and magnetic fields
which are responsible for non-perturbative dynamics.
His effective theory, however, also contains long-wavelength longitudinal
electric fields.
We discuss several subtleties in the treatment of longitudinal dynamics which
were not closely examined in B\"odeker's original treatment.
Somewhat to our surprise, we find that within its domain of validity
B\"odeker's effective theory does correctly describe both longitudinal and transverse
fluctuations.
We also show that, as far as the transverse dynamics of interest is concerned,
B\"odeker's effective theory could be replaced by a transverse-only theory 
that removes the longitudinal dynamics altogether.
In the process, we discuss several interesting aspects
of stochastic field theories.
}%
\end {abstract}

\thispagestyle{empty}


\section {Introduction}

Non-perturbative processes in a hot non-Abelian plasma
at or near equilibrium are associated with slow evolution of
magnetic gauge fields.%
\footnote
    {%
    This is explicitly argued in ref.\ \cite {non-perturb},
    but this fact is also implicit in
    earlier analysis of specific thermal effects such as plasmon damping rates
    of fast-moving particles \cite {damping-rates,smilga,BI2}
    and the color conductivity \cite {color-conductivity}.
    }
The characteristic spatial scale $R$ of non-perturbative gauge field
fluctuations
and the associated time scale $t$ for their evolution are of order
\begin {mathletters}
\label{eq:scales}
\begin {equation}
   R \sim {1 \over g^2 T},
   \qquad\qquad
   t \sim {1 \over g^4 T \ln(1/g)} ,
\label{eq:RTscale}
\end {equation}%
for small coupling.
Alternatively, the characteristic spatial momentum $k$
and frequency $\omega$ are
\begin {equation}
   k \sim g^2 T,
   \qquad\qquad
   \omega \sim g^4 T \ln(1/g).
\label{eq:kwscale}
\end {equation}%
\end{mathletters}%
For a review, see the introduction of our earlier paper \cite{paper1}.
The logarithm appearing in the time scale is a recent and interesting
result of B\"odeker \cite{Bodeker},
whose physical interpretation we discuss in ref.~\cite{paper1}.

Throughout this discussion, ``hot'' means that the temperature is large enough
that the running coupling $\alpha(T)$ is small,
that chemical potentials are ignorable, and that
there is no spontaneous symmetry breaking.
Examples of non-perturbative processes include chirality violation
in hot QCD, and baryon number violation in hot electroweak theory
(in its high-temperature symmetric phase).

B\"odeker \cite{Bodeker} has proposed an effective theory appropriate
for the scales (\ref{eq:scales}) above.
His effective theory is a {\it classical}\/ field
theory that involves only gauge fields with dynamics
governed by the diffusive Langevin equation
\begin {equation}
   \sigma \, \E = \D \times \B - \bnoise \,.
\label{eq:eff}
\end {equation}
Here, $\D$ is the covariant derivative acting in the adjoint representation.
In B\"odeker's proposal, $\bnoise$ is a Gaussian white noise random force,
normalized as%
\footnote
    {%
    We will scale our gauge fields by a factor of $g$,
    so covariant derivatives contain no explicit couplings
    while the action (or energy) has an overall factor of $1/g^2$.
    In addition, we will take the gauge group generators,
    and also the gauge field $\A \equiv \A^a \, T^a$,
    to be anti-Hermitian.
    Hence, the covariant derivative is simply $\D = \grad + \A$,
    and for the adjoint representation $(T^a)_{bc} = f_{bac}$
    with the structure constants $f_{abc}$ real and totally anti-symmetric.
    }
\begin {equation}
   \left\langle \noise^a_i(t,\x) \, \noise^b_j(t',\x') \right\rangle^{\strut}
     = 2 \sigma \, g^2 T \,\delta^{ab} \delta_{ij}
     \,\delta(t{-}t') \,\delta^{(3)}(\x{-}\x'),
\label{eq:corr}
\end {equation}
where $i,j$ and $a,b$ are spatial vector and adjoint color indices,
respectively.
This effective theory is supposed to give a quantitative description
of non-perturbative physics in the hot plasma to leading order in
the logarithm of the coupling.
In other words, corrections to this effective theory are
suppressed only by powers of $1/\ln (1/g)$.
In ref.\ \cite{paper1}, we showed that $\sigma$ can be interpreted as
the color conductivity%
\footnote{
   We are using ``color'' as a descriptive name for some non-Abelian
   gauge field.  It should be emphasized that all discussion of ``color''
   is applicable to the dynamics of, in particular,
   the SU(2) electroweak gauge field.
}
of the plasma, which is given by
\cite{color-conductivity,Bodeker,paper1}
\begin {equation}
   \sigma \approx {m_\pl^2 \over \gamma_\g} \,,
\end {equation}
where $m_\pl$ is the plasma frequency and
\begin {equation}
   \gamma_\g \approx \alpha \CA T \ln(1/g)
\label{eq:gammag}
\end {equation}
is the damping rate for hard thermal gauge bosons
\cite {damping-rates,smilga,BI2}.
The $\approx$ sign indicates equality
at leading logarithmic order.
[That is,  we are not distinguishing $\ln (2/g)$ from $\ln (1/g)$ in
Eq.~\ref {eq:gammag}, but the coefficient of the logarithm is correct.]
The plasma frequency $m_\pl$ is well known%
\footnote{
   For hot electroweak theory with a single Higgs doublet, for instance,
   $m_\pl^2 = {1\over18}(5+2\nf) \, g^2 T^2$ at leading order in $g$,
   where $\nf=3$ is
   the number of fermion families, and the
   adjoint Casimir $\CA=2$ in (\ref{eq:gammag}).
   For QCD with $n$ flavors of quarks,
   $m_\pl^2 = {1\over3}\left(1 + {n\over6}\right) g^2 T^2$
   and $\CA=3$, where $n$ is the number of relevant quark flavors
   (u, d, s, c, b, t).
}
at leading order in coupling
and is of order $gT$.

B\"odeker's effective theory is well suited to numerical simulation because
it is classical, insensitive to ultraviolet cut-offs \cite{paper1}, and
when cast into $A_0=0$ gauge generates a straightforward local equation
of motion for the evolution of $\A$:
\begin {equation}
   \sigma \, {d\over dt} \A = - \D\times\D\times\A + \bnoise .
\end {equation}
A numerical investigation of B\"odeker's effective theory and its
implication for electroweak baryon number violation has been recently
carried out by Moore \cite{Moore}.

Nevertheless, there is something peculiar about the effective theory
(\ref{eq:eff}).
In a high temperature plasma, static electric fields (or, more generally,
longitudinal fields) are Debye screened \cite {screening}.
The screening distance is of order $1/g T$,
which is small compared to the spatial scale $R \sim 1/g^2 T$ of
interest to us.  More generally, the longitudinal modes of the
gauge field are screened while, at low frequencies $\omega \ll k$,
the transverse modes are not.  Because of Debye screening,
it is the transverse electric and magnetic fields which are relevant
for producing non-perturbative fluctuations at the scales (\ref{eq:scales})
quoted earlier.
Longitudinal fields are irrelevant.
Nonetheless, B\"odeker's effective theory (\ref{eq:eff}) does describe
long-distance longitudinal fluctuations.
The longitudinal fields are those pieces
of $\E$ which contribute to $\D\cdot\E$ and which perturbatively
correspond to polarizations parallel to the momentum $\k$.
Dotting $\D$ into both sides of (\ref{eq:eff}), one sees that
\begin {equation}
   \sigma \, \D\cdot\E = - \D\cdot\bnoise .
\label{eq:DdotE}
\end {equation}
$\D\cdot\E$ is therefore not zero,
and so the fields in B\"odeker's effective theory
(\ref{eq:eff}) are not purely transverse.
Our purpose in this paper will be
to show two things: first, that B\"odeker's original derivation
of the long-distance longitudinal dynamics relied on a questionable approximation
which ignored subtleties associated with longitudinal dynamics,
but that the end result is nevertheless correct;
and, secondly, that the longitudinal dynamics is irrelevant
and may be removed altogether, if one is purely interested
in describing physical, gauge-invariant quantities that depend only on the
transverse fields (for example, the rate of anomalous charge violation).
The last point would be trivial in an Abelian theory,
because then (\ref{eq:eff}) would be linear in the fields and could be
projected into one equation involving only the transverse fields and another
independent equation involving only the longitudinal fields.  The point is
much less trivial in a non-Abelian theory because of the non-linearity of
(\ref{eq:eff}).


\section {Review of B\"odeker's derivation}

\subsection {The effective Boltzmann-Vlasov equation}

We will refer to gauge fields associated with the scales of interest
(\ref{eq:scales}) as ``soft'' fields.  In contrast, the dominant excitations
in the hot plasma correspond to momenta of order $T$ and will be
called ``hard.''
On his way to deriving the effective theory (\ref{eq:eff}) for the
soft gauge fields, B\"odeker \cite{Bodeker} first derived an effective
Boltzmann-Vlasov equation for the interaction of those fields with hard
excitations:
\begin {mathletters}%
\label{eq:vlasov}%
\begin {equation}
   (D_t + \v\cdot\D_\x) \, W - \E\cdot\v - \noiseA
   = - \delta C[W] ,
\label{eq:boltzW}
\end {equation}%
\begin {equation}
   D_\nu F^{\mu\nu} = J^\mu
    \equiv \mD^2 \left\langle v^\mu W(\v) \right\rangle_\v ,
\label{eq:JW}
\end {equation}%
\end {mathletters}
where
\begin {mathletters}%
\label{eq:dCW}%
\begin {equation}
   \delta C[W](\v) \equiv \gamma_\g
   \left\langle \I(\v,\v') \, W(\v') \right\rangle_{\v'} \,,
\end {equation}%
and
\begin {equation}
   \I(\v,\v') \equiv
      \deltaS(\v{-}\v')
      - {4\over\pi}
            {(\v\cdot\v')^2 \over \sqrt{1-(\v\cdot\v')^2}} \, .
\end {equation}
\end {mathletters}
Here, $\mD^2 \equiv 3 m_\pl^2$ is the leading-order Debye mass (squared).
The first equation (\ref{eq:boltzW})
is a linearized Boltzmann equation for the hard particles in the
presence of a soft electromagnetic field, where
$W(\v,\x,t)$ represents the color distribution of those particles and
$\v$ is a unit vector representing the hard particles' velocities.%
\footnote
    {%
    Technically,
    $W$ is the adjoint representation piece of the density matrix
    describing the color charges of the hard excitations,
    summed over the various species of excitations and
    integrated over the energy of excitations
    (for a fixed direction of motion $\v$).
    It is normalized in a way that simplifies the resulting
    equation.
    See ref.\ \cite{paper1} for the explicit definition.
    }
$\delta C$ represents a linearized collision term for $2{\to}2$ scattering
that randomizes the color charges of the hard particles \cite{paper1},
and $\noiseA$ is a source of random thermal noise.
The second equation (\ref{eq:JW}) is Maxwell's equation, where all the fields
on the left-hand side are to be understood as soft fields, and the
current on the right side is the soft-momentum component of the current
created by hard excitations.  This current is proportional to the
density $W$ of hard particles and the velocities of those particles,
where $v^\mu$ means $(1,\v)$.
In the explicit form (\ref{eq:dCW}) for the collision term,
$\langle\cdots\rangle_\v$ denotes averaging over the direction of $\v$
and $\deltaS$ is a $\delta$ function defined on the unit two-sphere with
normalization
\begin {equation}
   \left\langle \deltaS(\v{-}\v') f(\v') \right\rangle_{\v'} = f(\v) .
\end {equation}
See Refs.~\cite {Bodeker,paper1} for the derivation of the explicit
form (\ref {eq:dCW}b) of the linearized collision operator.

One may avoid worrying about the details of noise terms such
as $\noiseA$ until one reaches the final effective equation (\ref{eq:eff}),
at which point it is possible to then argue how the noise must in fact appear
\cite{paper1}.
However, since in this paper we will be discussing various subtleties,
it will be useful to keep track of the noise explicitly at each step
we consider.
In particular, B\"odeker \cite{Bodeker} derived that the appropriate
normalization of the noise in the effective Boltzmann-Vlasov equation
(\ref{eq:vlasov}) is related to the collision integral:
\begin {equation}
   \left\langle \noiseA^a(t,\x,\v) \, \noiseA^b(t',\x',\v') \right\rangle =
   { 2g^2T \over 3\sigma} \, \I(\v,\v') \,
   \delta^{ab}\,\delta(t{-}t') \, \delta^{(3)}(\x{-}\x') \,.
\label{eq:corrA}
\end {equation}
We will not review any further the origin of the effective Boltzmann-Vlasov
equations (\ref{eq:vlasov}) and direct the reader instead to B\"odeker's
original work \cite{Bodeker} and our alternative derivation
\cite{paper1}.  It is in the step from these kinetic equations to
B\"odeker's final effective theory (\ref{eq:eff}) that subtleties in
the treatment of longitudinal physics creep in, and that is the focus
of this paper.


\subsection {Solving for $W$}
\label{sec:naiveW}

B\"odeker obtains his final effective theory (\ref{eq:eff}),
at leading-log order, from the
effective Boltzmann-Vlasov equations (\ref{eq:vlasov}) by arguing that
the covariant derivative terms in the Boltzmann equation
(\ref{eq:boltzW}) are together of order $g^2 T \, W$
and so can be ignored compared to the collision term,
which is of order $\gamma_\g W \sim (g^2 T \ln g^{-1})\,W$ and hence
larger by a logarithm.
There is an important subtlety to this approximation which will be examined
in the next section.
But accepting this argument at face value for now,
if one drops the covariant derivative terms then the
Boltzmann equation becomes simply
\begin {equation}
   \E\cdot\v + \noiseA
   \approx \delta C[W] .
\label{eq:dropD}
\end {equation}%
Formally, the solution is
\begin {equation}
   W = (\delta C)^{-1} (\E\cdot\v + \noiseA) ,
\end {equation}
where $\delta C$ is to be understood here as an operator acting on the space
of (adjoint-representation) functions of a unit vector $\v$.
This result for $W$ yields the spatial current appearing in (\ref{eq:JW}),
\begin {equation}
   \J = \mD^2 \left\langle \v \, (\delta C)^{-1}
            (\E\cdot\v + \noiseA) \right\rangle_\v .
\label{eq:J1}
\end {equation}

Next note that $\delta C$ preserves the parity (in $\v$)
of functions it acts on.
In other words, $\delta C$ maps even (odd) functions
of $\v$ into even (odd) functions of $\v$.
(In contrast, the $\v\cdot\D_\x$ operator that we dropped does not.)
Moreover, in the space of odd functions of $\v$, $\delta C$ as given
by (\ref{eq:dCW}) reduces to simply $\delta C = \gamma_\g$.
So, since $\delta C$ is a symmetric operator, and since it acts to the left
on the odd function $\v$ in (\ref{eq:J1}), we can replace $(\delta C)^{-1}$
by $\gamma_\g^{-1}$ in that equation to obtain
\begin {equation}
   \J = {\mD^2 \over \gamma_\g}
	\left\langle \v \, (\E\cdot\v + \noiseA) \right\rangle_\v
      = \sigma \E + \bnoise ,
\label{eq:JE}
\end {equation}
where
\begin {equation}
   \bnoise \equiv 3\sigma \langle \v \, \noiseA \rangle_\v .
\end {equation}
Using the correlation (\ref{eq:corrA}) for $\noiseA$, one produces the
correlation (\ref{eq:corr}) asserted earlier for $\bnoise$.
Taking the spatial part of the Maxwell equations (\ref{eq:JW}) and
dropping the $d\E/dt$ term which,
for the scales (\ref{eq:scales}) of interest,
is smaller (by four powers of coupling) than the $\sigma \, \E$ term,
one obtains B\"odeker's final effective theory (\ref{eq:eff}).


\section {Longitudinal subtleties}

In the introduction, we noted that B\"odeker's effective theory
(\ref{eq:eff}) contains a fluctuating longitudinal electric field.
This may seem puzzling since longitudinal electric fields are Debye screened.
In this section, we will take a closer look at how both
Debye screening, and B\"odeker's effective Langevin equation,
do emerge from the Boltzmann-Vlasov equations~(\ref {eq:vlasov}).


\subsection {Zero mode of $\delta C$}
\label{sec:zmode}

In the last section, the Boltzmann equation for $W$ was simplified,
at leading-log order,
by arguing that $\delta C$ dominates over the convective derivative
$D_t + \v\cdot\D_\x$ by a power of $\ln(g^{-1})$.  This is not quite
correct, however, because the operator $\delta C$ has an eigenvalue which
is {\it not} order $\gamma_\g$ and which does not dominate over the
convective derivative; specifically, $\delta C$ has a zero mode.

The necessity of this zero mode was noted by B\"odeker, who observed that
the effective Maxwell equation (\ref{eq:JW}) for the soft fields requires
conservation $D_\mu J^\mu = 0$ of the current
$J^\mu = \mD^2 \langle v^\mu W(\v) \rangle_\v$ generated by
the hard particles.  From (\ref{eq:boltzW}), this conservation requires
$\langle \delta C[W] \rangle_\v = 0$, which is indeed satisfied by
(\ref{eq:dCW}).

The fact $\langle \delta C[W] \rangle_\v = 0$ can be rephrased to
say that the symmetric operator $\delta C$ has null states: it
annihilates anything that is independent of $\v$.
(
This can be written in bra-ket notation in $\v$-space as
$\langle \hbox{constant} | \delta C | W \rangle =
\langle W | \delta C | \hbox{constant} \rangle = 0$ for any $W$.)
This point will be important later on,
so let us give an alternative way of understanding it.
The collision term $\delta C$ does not care, at leading-log order, about the
dynamics of the soft fields.  In particular, it does not care that the soft
effective theory is a gauge theory, with a local color symmetry, instead of a
non-gauge theory, with merely a global color symmetry.
So, from the point of view of the calculation of $\delta C$ at leading-log
order, the theory {\it could\/} have been one where it was meaningful
to talk about the total color charge of the system.  
If one then imagined adding an infinitesimal chemical potential $\mu$
for this total color charge, the resulting equilibrium density would be
\begin {equation}
   n = \left[e^{\beta (\epsilon_\p - g \mu_a T^a)} \mp 1\right]^{-1}
      = n_0 + n_0 (1 {\pm} n_0) \, \beta g \, \mu_a T^a + O(\mu^2)
\label {eq:nmu}
\end {equation}
for each particle type, where $n_0$ is the $\mu=0$ equilibrium distribution.
In equilibrium, the collision term in a Boltzmann equation always vanishes
by detailed balance.
Different values of $\mu$ correspond to different equilibrium states,
and the collision term must therefore vanish for all $\mu$.
That means that the linearized deviation
\begin {equation}
   \delta n =  n_0 (1 {\pm} n_0) \, \beta g \mu_a T^a
\label{eq:dn}
\end {equation}
of the equilibrium distribution (\ref{eq:nmu}) away from $n_0$
must correspond to a null state of the linearized collision
operator $\delta C$.
The deviation (\ref{eq:dn}) is isotropic and homogeneous---it is independent
of both $\v$ and $\x$.
As a result, when re-expressed in terms of the function $W(\x,\v)$
used to parametrize color distributions of the hard particles in
the linearized Boltzmann equation (\ref{eq:boltzW}),
the deviation (\ref{eq:dn})
corresponds to $W(\x,\v) =$ constant.
That means that a constant $W$ is a null vector of $\delta C$.
But since collisions are local in $\x$ (in the effective theory),
the $\x$ dependence of $W$ is irrelevant, and so any $W$ which
does not depend on $\v$ is a null vector of the linearized collision
operator $\delta C$.


\subsection {Longitudinal and transverse projections}

Before continuing, it is worthwhile to introduce longitudinal and
transverse projection operators.  Perturbatively, the longitudinal
projection operator for the electric field is
\begin {equation}
   (\PLpert)^{ij} =
   \hat k^i \hat k^j =
   \nabla^i {1 \over \nabla^2} \, \nabla^j \,.
\end {equation}
The gauge-covariant non-perturbative generalization is
\begin {equation}
   \PL^{ij} \equiv D^i {1\over D^2} \, D^j ,
\label{eq:PLdef}
\end {equation}
where $D^2$ means $\D \cdot \D$.
The transverse projection operator is of course
\begin {equation}
   \PT^{ij} = \delta^{ij} - \PL^{ij} .
\end {equation}
It is the longitudinal electric field which couples to
external charges, since Gauss' Law reads $\D\cdot\E = \rho$ and
since $\D\cdot(\PT\,\E) = 0$.  And it's the transverse
electric field that is produced by $\D\times\B$ in the effective
theory (\ref{eq:eff}) since $\PL \, (\D\times\B) = 0$.
As mentioned in the introduction,
the precise separation between longitudinal
and transverse dynamics is not transparent from this
simple discussion because of the non-linear dependence of
$\D\times\B$ on the underlying vector potential $\A$.


\subsection {Solving for $W$ (again)}

To examine the difficulties caused by the presence of a
zero mode of $\delta C$,
we now return to the effective Boltzmann-Vlasov equation (\ref {eq:vlasov})
and will repeat the analysis of section \ref{sec:naiveW},
this time being more careful about how we treat the convective
derivative compared to $\delta C$.
Formally, the solution for $W$ is
\begin {equation}
   W = {1 \over D_t + \v\cdot\D_\x + \delta C} \> (\v\cdot\E + \noiseA) ,
\label {eq:Weq}
\end {equation}
or $W = G (\v\cdot\E + \noiseA)$, where $G$ denotes the
inverse of the linearized kinetic operator,
\begin {equation}
    G \equiv \left[ D_t + \v\cdot\D_\x + \delta C \right]^{-1} \,.
\label {eq:G}
\end {equation}
This fluctuation in the distribution of hard excitations
produces a current response [from Eq.~(\ref {eq:JW})] of
\begin {equation}
   \J = \mD^2
       \left\langle \v \, G \, (\v\cdot\E + \noiseA) \right\rangle_\v ,
\label{eq:JL0}
\end {equation}
and the (color) charge density
\begin {equation}
    \D \cdot \E = J^0 = \mD^2
   \left\langle G \, (\v\cdot\E + \noiseA) \right\rangle_\v .
\label {eq:J0}
\end {equation}
One may easily check 
that the current is conserved (as it must be), since
\begin {eqnarray}
    D_0 J^0 + \D \cdot \J
    &=&
    \mD^2
    \left\langle
	(D_t + \v\cdot\D) \, G \, (\v\cdot\E + \noiseA)
    \right\rangle
\nonumber\\ &=&
    \mD^2
    \left\langle
	\left[ 1 - \delta C \, G \right]
	(\v\cdot\E + \noiseA)
    \right\rangle
\nonumber\\ &=&
    0 \,.
\end {eqnarray}
The $\langle \v\cdot\E + \noiseA \rangle_\v$ term vanishes due to isotropy,
$\langle \v \rangle_\v = 0$, and the lack of bias in the noise,
$\langle \noiseA \rangle_\v = 0$.
And $\langle \delta C \, G (\v\cdot\E + \noiseA) \rangle_\v$
vanishes because $\delta C$ is acting (to the left) on its
$\v$-independent zero-mode.%
\footnote
    {%
    In slightly more explicit notation, this term is
    $
	\gamma_\g 
	\left\langle
	\I(\v'',\v') \, G(\v',\v) \,
	[\v\cdot\E + \noiseA(\x,\v)]
	\right\rangle_{\v,\v',\v''}
    $.
    It vanishes because
    $
	\left\langle \I(\v'',\v') \right\rangle_{\v''} =
	\left\langle \I(\v'',\v') \right\rangle_{\v'} = 0
    $.
    }

\subsection {The problem with the naive derivation}

In following subsections, we will discuss how to evaluate the operator
inverse that defines $G$.
To begin, however, it is useful to see how the zero-mode problem manifests
itself in a simple example of matrix inversion.
To this end, let us for the moment
replace the Green function $G$ of (\ref{eq:G}) by
that of a simplified finite dimensional example.
First, imagine that the gauge interactions are Abelian, so that
$D_t$ and $\D_\x$ can be replaced by simply $-i \omega$ and $i\k$,
respectively.
Next, imagine that the infinite-dimensional space of possible
functions of $\v$, on which 
$\delta C$ acts, is
truncated to the four-dimensional space of functions that
are either independent of $\v$ or linear in $\v$.
We wish to examine the matrix representing the action of
$-i \omega + \v\cdot\D_\x + \delta C$ within this truncated space.
In order to distinguish clearly between longitudinal and
transverse physics, it is convenient to choose a basis
$\{ f_\alpha (\v) \}$, $\alpha = 0,\ldots,3$, where
\begin {equation}
   f_0(\v) = 1 , \qquad
   f_i(\v) = \sqrt 3 \, \hat e_{i} \cdot \v \,,
\end {equation}
and $\hat e_{i}$ are three mutually orthonormal unit vectors with
$\hat e_1 \equiv \hat k$ pointing in the direction of $\k$.
The overall normalization has been chosen so that
$\langle f_i f_j \rangle_{\v} = \delta_{ij}$.
In this basis, the matrix elements of 
$\langle f_i | {-}i\omega + \v\cdot\D_\x + \delta C | f_j \rangle_{\v}$
are
\begin {equation}
   \left( \begin {array}{cccc}
        -i\omega         &  {i\over\sqrt3} \, k &   & \cr
        {i\over \sqrt3} \, k  &  \gamma_\g        &   & \cr
                         &                  & -i\omega {+} \gamma_\g & \cr
                         &                  &   & -i\omega {+} \gamma_\g \cr
   \end {array}\right) .
\end {equation}
The inverse operator, corresponding to $G$ in our truncated space, is
\begin {equation}
   G_{\rm trunc} = 
   \left( \begin {array}{rrrr}
    \gamma_\g / (-i\gamma_\g\omega {+} {1\over3} k^2)
	& -{i\over \sqrt3} \, k / (-i\gamma_\g\omega {+} {1\over3} k^2)
	& & \cr
    -{i\over \sqrt3} \, k / (-i\gamma_\g\omega {+} {1\over3} k^2)
	& -i\omega / (-i\gamma_\g\omega {+} {1\over3} k^2)
	& & \cr
	& & (-i\omega {+} \gamma_\g)^{-1} & \cr
	& & & ( -i\omega {+} \gamma_\g)^{-1} \cr
   \end {array} \right) .
\label {eq:ourm0}
\end {equation}
The $\omega\to 0$ limit is particularly simple:
\begin {equation}
   G_{\rm trunc} \to 
   \pmatrix{
        \phantom- {3\>\gamma_\g \over k^2} &
             -{i\sqrt3 \over k} &   & \cr
        -{i\sqrt3 \over k} &
             0   &   & \cr
                         &                  & \gamma_\g^{-1} & \cr
                         &                  & & \gamma_\g^{-1} \cr
   } .
\label{eq:ourm}
\end {equation}
In contrast, the naive derivation of B\"odeker's theory corresponds
to replacing $-i\omega + i\v\cdot\k + \delta C$ by $\delta C$, and
the corresponding ``inverse'' would then be
\begin {equation}
   \pmatrix{
        \infty & &   & \cr
        & \gamma_\g^{-1} &   & \cr
                         &                  & \gamma_\g^{-1} & \cr
                         &                  & & \gamma_\g^{-1} \cr
   } .
\label {eq:naivem}
\end {equation}

As one can see, there is no difference in the transverse sector
(spanned by $f_{2,3}$), but there is a huge difference in the
longitudinal sector.
As a particular example, consider the noiseless
part of
the current $\J$, given in Eq.~\ref{eq:JL0};
namely $\md^2 \langle \v \, G \, \v\cdot\E\rangle_\v$.
In the naive derivation, as represented by (\ref{eq:naivem}),
this contribution gives
\begin {equation}
   \md^2 \langle \v \, \gamma_\g^{-1} \, \v\cdot\E\rangle_\v
   = {\md^2\over3\gamma_\g} \, \E .
\end {equation}
In our Abelian truncated-space calculation of $G$, however, it is instead
given by
\begin {equation}
   {\md^2\over3} (G_{11} \PL + G_{22} \PT) \E
   = {\md^2\over3\gamma_\g} \, \PT \E .
\end {equation}
in the $\omega\to0$ limit.
The longitudinal part of $\E$ is projected out!
This is quite different from the result of the naive derivation.


\subsection {Low-frequency, long-wavelength dynamics}

We will now show that,
despite the major difference in $G^{-1}$,
one nonetheless
{\it does}\/ recover B\"odeker's effective theory even for longitudinal
physics.
To do so, we will also return to the full, original non-Abelian
problem and dispense with the truncated Abelian model of the previous section.

If we restrict our attention to frequencies and wavenumbers which
are small compared to the damping rate, $\omega, k \ll \gamma_\g$,
then in the Greens' function $G$ we could drop the 
convective derivative compared to $\delta C$
were it not for the fact that $\delta C$ has a zero-mode.
To deal with this, let $P_0$ denote the projection operator
onto the zero-mode of $\delta C$, so that
\begin {equation}
    P_0 \left(f(\v)\right) \equiv \langle f(\v) \rangle_{\v} \,,
\end {equation}
and separate the convective derivative into zero-mode and
non-zero-mode pieces,
\begin {eqnarray}
    D_t + \v \cdot \D
    &=&
    \left( D_t + \v \cdot \D \right) P_0 + P_0 \, \v\cdot\D
    + (1{-}P_0) \left( D_t + \v \cdot \D \right) (1{-}P_0) \,.
\label {eq:convective}
\end {eqnarray}
To see this, note that $D_t$ commutes with $P_0$, and that
$P_0 \, \v\cdot\D \, P_0 = 0$.
The last term of (\ref{eq:convective}) only perturbs
the non-zero eigenvalues of $\delta C$, and
may be neglected provided $\omega$ and $k$ are small compared
to $\gamma_\g$.
[For $k = O(g^2 T)$ this is a leading-log approximation.]
The first two terms of (\ref {eq:convective}) are rank one
perturbations which will lift the zero-mode of the linearized
kinetic operator.
And one may evaluate explicitly the change in the inverse of
an operator produced by adding a finite rank perturbation.
In this case, a short exercise shows that
\begin {eqnarray}
    G
    &\simeq&
    \left[ (D_t + \v\cdot \D) P_0 + P_0 \, \v\cdot\D + \delta C \right]^{-1}
\nonumber \\&=&
    (1{-}P_0) \, \delta C^{-1} (1{-}P_0)
    +
    \gamma_\g^{-1}
    (\gamma_\g - \v\cdot\D) \,
    {P_0 \over \gamma_\g D_t - {1\over 3} \D^2} \,
    (\gamma_\g - \v\cdot\D) \,.
\label {eq:Gapprox}
\end {eqnarray}
(To verify the last equality, recall that $\delta C$ is nothing but
multiplication by $\gamma_\g$ when acting on odd functions of $\v$.
Hence, $\delta C \, \v\cdot\D \, P_0 = \gamma_\g \, \v\cdot\D \, P_0$.)

Now pause to note the correspondences of this result with
the truncated Abelian results of the previous section.  The
full inverse (\ref{eq:Gapprox}) gives
\begin {mathletters}
\begin {eqnarray}
   \langle G \rangle_\v
   &\simeq& \gamma_\g \left[\gamma_\g D_t - \third \D^2\right]^{-1} ,
\label{eq:Gvevapprox}
\\
   \langle \v G \rangle_\v
   &\simeq& {\third} \D \left[\gamma_\g D_t - \third \D^2\right]^{-1} ,
\\
   \langle G \v \rangle_\v
   &\simeq& \left[\gamma_\g D_t - \third \D^2\right]^{-1} {\third} \D ,
\\
   \langle \v G \v \rangle_\v
   &\simeq& \gamma_\g \left\{ 1 -
     {\third} \D \left[\gamma_\g D_t - \third \D^2\right]^{-1} {\third} \D
   \right\}
   \quad \mathop{\longrightarrow}\limits_{D_t \to 0} \quad
   {1\over3\gamma_\g} \PT \,,
\end {eqnarray}%
\end {mathletters}
for $\omega, k \ll \gamma_\g$.
These are simple non-Abelian generalizations of (\ref{eq:ourm0}).
(The factor of 3 differences just reflect the normalizations of
our chosen basis functions in the last section.)

The form (\ref{eq:Gapprox})
for $G$ may now be inserted into Eqs.~(\ref {eq:JL0})
and (\ref {eq:J0}).
For the charge density, one finds
\begin {equation}
    \D\cdot\E = J^0 =
    - \gamma_\g
    \left[ \gamma_\g D_t - {\textstyle {1\over3}} \D^2 \right]^{-1}
    \left( \sigma \D\cdot\E + \D\cdot\bnoise \right) ,
\label{eq:DE1}
\end {equation}
or
\begin {equation}
    \left[
	D_t + {{\sigma\over\mD^2}} \left(-\D^2 + \mD^2 \right)
    \right]
    \D\cdot\E
    =
    -\D\cdot\bnoise \,,
\end {equation}
where $\sigma \equiv {\textstyle {1\over3}} \mD^2 / \gamma_\g$,
and $\bnoise \equiv 3\sigma \langle \v \noiseA \rangle_{\v}$.
And for the current,
from (\ref{eq:JL0}) and (\ref{eq:DE1}),
\begin {equation}
    \J =
    \sigma \E + \bnoise - {\sigma \over \mD^2} \, \D
    \left( \D\cdot\E \right) .
\end {equation}
Inserting this into the Maxwell equation
$-D_t \, \E + \D \times \B = \J$ gives
\begin {equation}
    D_t \, \E + \sigma \E - {\sigma \over \mD^2} \, \D (\D \cdot \E)
    =
    \D \times \B - \bnoise \,.
\label {eq:eff1}
\end {equation}
This local equation of motion is the exact result which follows
from approximating $G$ as shown in (\ref {eq:Gapprox}).
However, that form for $G$ was based on the assumption that
the frequencies and wavevectors of interest are small compared
to the damping rate $\gamma_\g$.
Since $\gamma_\g$ is $O(g^2 T \ln g^{-1})$, this means
that $\omega$ is tiny compared to the $O(T/\ln g^{-1})$
conductivity, and that $k$ is much smaller than the $O(gT)$ Debye mass.
Hence, there is no point in retaining the $D_t \, \E$ or
$\D (\D\cdot\E)$ terms in the effective equation (\ref {eq:eff1}).
Dropping these terms immediately yields B\"odeker's equation
(\ref {eq:eff}).
In other words, a more careful treatment of the effect of the
zero mode in $\delta C$ does not produce any difference (in
leading-logarithmic approximation) to the resulting effective theory.

\subsection {Recovering Debye screening}

To see how Debye screening emerges from the kinetic theory
(\ref {eq:vlasov}),
return to Eq.~(\ref {eq:Weq}) for $W$ and now assume
that the scales of interest are in the perturbative regime
where $k \gg g^2T$ and/or $\omega \gg g^4 T \ln g^{-1}$.
[In other words, the necessary conditions (\ref {eq:kwscale})
for non-perturbative fluctuations are not both satisfied.]
In this regime, the gauge fields in the covariant derivatives
appearing in the Greens' function $G$ may be treated as small.%
\footnote
    {%
    For a more detailed justification, based on a computation
    of the power spectrum of gauge field fluctuations,
    see Ref.~\cite {ASY1}.
    }
Expanding $G$ in powers of the gauge field,
the leading term,
\begin {equation}
    G \simeq
    \left[ \, \partial_t + \v \cdot \nabla_\x + \delta C \right]^{-1} \,,
\end {equation}
is diagonal in momentum space.
Fourier transforming Eq.~(\ref {eq:J0}) for the charge density
then yields (to leading order in the gauge field)
\begin {eqnarray}
    i \, \k \cdot \E
    &=&
    \mD^2 \, \langle \tilde G \, \v \rangle_\v \cdot \E
    +
    \mD^2 \, \langle \tilde G \noiseA \rangle_\v 
\nonumber\\ &=&
    \mD^2 \, \langle \tilde G \, \v \cdot \k \rangle_\v {\k \cdot \E \over \k^2}
    +
    \mD^2 \, \langle \tilde G \noiseA \rangle_\v 
\nonumber\\ &=&
    -i \mD^2 \,
    \left( 1 + i \omega \langle \tilde G \rangle_\v \right)
    {\k \cdot \E \over \k^2} +
    \mD^2 \, \langle \tilde G \noiseA \rangle_\v  \,,
\label {eq:kE}
\end {eqnarray}
or
\begin {equation}
    \left[
	\k^2 + \mD^2 \left( 1 + i \omega \langle \tilde G \rangle_\v \right)
    \right]
    i \, \k \cdot \E
    =
    \mD^2 \, \k^2 \, \langle \tilde G \noiseA \rangle_\v  \,,
\label {eq:kEeqn}
\end {equation}
with
\begin {equation}
    \tilde G(\omega,\k) =
    \left [ -i \omega + i \v \cdot \k + \delta C \right]^{-1} \,,
\end {equation}
and $\E$ and $\noiseA$ now denoting the $(\omega,\k)$ Fourier components
of these fields.
In the first step of (\ref {eq:kE}), we used the fact that, with gauge
fields neglected in $\tilde G$,
the only vector which $\langle G \, \v \rangle_\v$
can depend upon is $\k$, and therefore only the longitudinal component
of $\E$ can contribute to the result.
The following step used
$
    \langle \tilde G \, (i\v\cdot\k -i\omega) \rangle
    =
    1 - \langle \tilde G \, \delta C \rangle
    =
    1
$
which is another consequence of the zero mode in $\delta C$.
The result (\ref {eq:kEeqn}) shows that $\k \cdot \E$ satisfies
a diffusive Langevin equation in which the noise and damping
depend on $\langle \tilde G \noiseA \rangle_\v$
and $\langle \tilde G \rangle_\v$,
respectively.

The power spectrum of charge density (or $\D \cdot \E$) fluctuations
is defined as
\begin {eqnarray}
    \rho_{\rm L}(\omega,\k)
    &\equiv&
    \int dt \, d^3\x \> e^{i\omega t -i \k \cdot \x} \,
    \langle J^0(t,\x) J^0(0,0) \rangle_\noiseA \,.
\end {eqnarray}
Using (\ref {eq:kEeqn}) to express $\D \cdot \E$ in terms of the noise
$\noiseA$,
and then recalling that the covariance of the noise $\noiseA$,
as given by (\ref {eq:corrA}) and (\ref {eq:dCW}),
is proportional to $\delta C$, allows one to write the power
spectrum as
\begin {eqnarray}
    \rho_{\rm L}(\omega,\k)
    &=&
    {\mD^4 \, \k^4
    \langle
	\langle \tilde G\noiseA \rangle_\v
	\langle \noiseA \tilde G^\dagger \rangle_{\v'}
    \rangle_\noiseA
    \Bigm/
	\left|\,
	    \k^2 + \mD^2 \left(1 + i \omega \langle \tilde G \rangle_\v \right)
	\right|^2
    }
\nonumber\\ &=&
    {2 g^2 T \, \mD^2 \, \k^4 \, {\rm Re}\, \langle \tilde G \rangle_\v
    \Bigm/
	\left|\,
	    \k^2 + \mD^2 \left(1 + i \omega \langle \tilde G \rangle_\v \right)
	\right|^2
    }
\nonumber\\ &=&
    -{2T \over \omega} \>
    {\rm Im} \>
    {g^2 \, \k^4
    \over
    \k^2 + \mD^2 \left(1 + i \omega \langle \tilde G \rangle_\v \right) }
    \,.
\label {eq:rhoL}
\end {eqnarray}
Once again,
this answer is valid provided $k \gg g^2 T$ and/or
$\omega \gg g^4 T \ln g^{-1}$ since we neglected (for this discussion only)
the soft gauge fields in the covariant derivatives.
Furthermore, $k$ and $\omega$ must be small compared to $T$,
since this is a basic requirement for any kinetic theory
description to be valid.

Linear response theory, applied to the underlying quantum
field theory, shows that the power spectrum
\begin {equation}
    \rho_{\rm L}(\omega,\k)
    \equiv
    \int dt \, d^3\x \> e^{i\omega t -i \k \cdot \x} \,
    \left\langle \half \left\{ J^0(t,\x) ,  J^0(0,0) \right\} \right\rangle \,,
\end {equation}
(defined with a symmetrized ordering of quantum operators)
is related to the retarded charge-density charge-density correlator
\begin {equation}
    D_{\rm R}(\omega,\k)
    \equiv
    i
    \int dt \, d^3\x \> e^{i\omega t -i \k \cdot \x} \>
    \theta(t)
    \left\langle \left[ J^0(t,\x) ,  J^0(0,0) \right] \right\rangle \,,
\label {eq:LRT}
\end {equation}
by the fluctuation-dissipation relation
\begin {equation}
    \rho_{\rm L}(\omega,\k)
    =
    [2n(\omega) + 1] \> {\rm Im} \, D_{\rm R} (\omega,\k) \,,
\end {equation}
where $n(\omega) = [e^{\beta \omega} {-}1]^{-1}$ is the equilibrium
Bose distribution function.
And the retarded correlator $D_{\rm R}(\omega,\k)$ is the analytic
continuation of the Euclidean space (time-ordered imaginary-time)
correlator $D_{\rm E}(i\omega_n,\k)$ from the imaginary Matsubara 
frequencies to (just above) the real frequency axis.

By comparing (\ref {eq:rhoL}) to the form (\ref {eq:LRT}),
the kinetic theory result for the power spectrum may be
converted to an equivalent result for the retarded correlator.
The leading factor of $T/\omega$ is just
the low-frequency (classical) limit of the Bose distribution
function, and hence the retarded charge density correlator is
\begin {eqnarray}
    D_{\rm R}(\omega, \k)
    &=&
    g^2 \, \k^2
    -
    {g^2 \, \k^4 \over
    \k^2 + \mD^2 \left(1 + i \omega \langle \tilde G\rangle_\v \right) } \,.
\label {eq:DR}
\end {eqnarray}
The local (and temperature-independent) $g^2 \, \k^2$ term,
which does not contribute to the imaginary part,
is determined by the current-current Ward identities,
or equivalently by the requirement that $D_{\rm R}$ remain bounded
as $\k \to \infty$.
The low frequency limit,
\begin {eqnarray}
    D_{\rm R}(0, \k)
    &=&
    {g^2 \, \k^2 \, \mD^2 \over \k^2 + \mD^2}
    =
    g^2 \left[ \mD^2 - {\mD^4 \over \k^2 + \mD^2} \right],
\end {eqnarray}
reproduces the correct static equilibrium Debye-screened
charge density correlations.%
\footnote
    {%
    The factor of $g^2$ is present merely because we have chosen to
    scale all our gauge fields by $g$ relative to the usual
    perturbative conventions.
    In, for example, Coulomb gauge,
    $D_{\rm R}$ equals the one-loop gauge field
    self-energy $\Pi^{00}$ (which is just $\mD^2$ in the static limit),
    plus the one-particle reducible contributions which sum to
    $\Pi^{00} D^{00} \Pi^{00}$,
    where $D^{00} = \langle A^0 A^0 \rangle = -1 / (\k^2 + \Pi^{00})$
    is the Debye-screened $A^0$ propagator.
    }
More generally, the kinetic theory answer (\ref {eq:DR})
reduces to the known hard-thermal-loop result
whenever the frequency or momentum is large compared to the
damping rate $\gamma_\g$.
In this domain, the details of the scattering operator $\delta C$
are irrelevant, and $\tilde G(\omega,\k)$ may be approximated by
$
    i [\omega - \v \cdot \k + i \epsilon]^{-1}
$.
The resulting average over $\v$ may then be performed analytically,
and yields the standard HTL result for the self-energy.

If neither $k$ nor $\omega$ are large compared to the damping
rate $\gamma_\g$, then the detailed form of $\delta C$ is significant.
Evaluating $\tilde G(\omega,k)$ is non-trivial
if $k$ is comparable to the damping rate.
However, if $k$ and $\omega$ are both small compared to $\gamma_\g$,
then the previous representation (\ref {eq:Gvevapprox})
may be used.  Perturbatively, it gives
\begin {equation}
    \langle \tilde G (\omega,\k) \rangle_{\v}
    =
    {\gamma_\g \over {1\over3} \k^2 -i \omega \gamma_\g } \,,
\end {equation}
which gives
\begin {equation}
    \left. \rho_{\rm L}(\omega,\k) \right|_{\hbox{\tiny B\"odeker}}
    =
    {2 g^2 T \over \sigma} \, \k^2 \,.
\end {equation}
This is the same result which emerges directly from the relation
$\sigma\D\cdot\E = -\D\cdot\bnoise$ in B\"odeker's effective theory,
combined with (\ref{eq:corr}) for the noise variance.
This result is valid in the overlap of the perturbative and
damping-dominated domains, that is when $|k^2 -i \omega \sigma| \gg g^4 T^2$
and $|{1\over3}k^2 -i \omega \gamma_\g| \ll \gamma_\g^2$.


\section {Irrelevancy of Longitudinal Dynamics}

We have seen that
B\"odeker's effective theory
\begin {equation}
   \sigma \E = \D \times \B - \bnoise \,,
\label{eq:bodeker2a}
\end {equation}
within its domain of validity,
does correctly describe both longitudinal and transverse fluctuations.
However, it is the transverse part of the gauge fields which are responsible
for interesting non-perturbative phenomena such as topological
transitions and associated baryon non-conservation.
One may wonder if it is possible to formulate an equally valid
effective theory which describes {\em only} transverse physics.
On the face of things, this should be easy;
just insert a transverse projection operator
to eliminate the longitudinal part of the noise,
\begin {equation}
   \sigma \E = \D \times \B - \PT \, \bnoise \,.
\label{eq:mea}
\end {equation}
This produces an effective theory with no longitudinal dynamics
whatsoever, $\PL \E = 0$.
In the case of an Abelian theory, the trivial decoupling of transverse and
longitudinal parts of the gauge field would make it obvious that
B\"odeker's theory (\ref {eq:bodeker2a})
and the transverse-projected theory (\ref {eq:mea})
describe exactly the same transverse dynamics.
But for our non-Abelian theory, the dependence
of covariant derivatives and projection operators on the gauge field
makes this decoupling far less obvious.
The remainder of this paper is devoted to showing that it is {\it almost}\/
true that equations (\ref {eq:bodeker2a}) and (\ref {eq:mea})
do in fact generate identical transverse dynamics.
The ``almost'' caveat reflects subtleties associated with the fact
that white noise cannot be considered smooth
even on infinitesimal time scales.
Investigating these subtleties will
reveal that the naive transverse equation (\ref{eq:mea})
must be corrected but can then be made exactly equivalent to
B\"odeker's equation (\ref{eq:bodeker2a}) with regard to the
transverse dynamics.
It should be emphasized that this investigation was motivated
by theoretical curiosity, not practical convenience.%
\footnote
    {
    In fact, historically, it took us some time to recognize that
    longitudinal physics for $\omega \ll k \ll \gamma_\g$ is
    correctly reproduced by B\"odeker's effective theory,
    and the discussion in this section was motivated by the desire
    to show that it simply doesn't matter.
    }
B\"odeker's equation (\ref {eq:bodeker2a}) is local,
whereas the transverse-projected equation (\ref {eq:mea}) is not.
Consequently, numerical simulations are far more easily performed
in the original theory than in any transverse-projected variant.

To be precise, by ``transverse dynamics'' we refer to all
physical observables that do not depend on $A_0$ when
expressed as gauge-invariant functions of the fields $A_\mu$.
The magnetic field $\B = \D\times\A$ does not, of course,
depend on $A_0$.  Writing $\E = \D A_0 - (d\A/dt)$, it is easy to see
that the transverse electric field $\PT \, \E = -\PT \, (d\A/dt)$ does not
either.
Consequently, an example of a physical quantity which depends
only on the transverse dynamics is the topological charge
(or change in Chern-Simons number) of the gauge field,
which is proportional%
\footnote{
   The precise formula is
   $\Delta N(t) = (1/8\pi^2) \int_0^t dt \> d^3x \> E_i^a B_i^a$.
}
to $\int\tr[\E\cdot\B] = \int\tr[(\PT\E)\cdot\B]$.%
\footnote
{%
The topological transition rate (or Chern-Simons number diffusion constant),
is an important ingredient in scenarios of electroweak baryogenesis.
Understanding the applicability of numerical simulations
using B\"odeker's effective theory for extracting the topological
transition rate motivated this investigation.
See \cite {Moore} for such recent numerical work and related discussion.
}


\subsection {Naive equivalence}
\label{sec:naive}

We first wish to paint with a broad brush.
We will for the moment ignore all
subtleties and discuss how,
if one implicitly and incorrectly (and only when advantageous) treats the
noise $\bnoise(\x,t)$ as a smooth function of $t$, one may show that
the two theories (\ref{eq:bodeker2a}) and (\ref{eq:mea}) should generate
the same transverse dynamics.
We will wait until section \ref{sec:wrong} and its sequel
to correct this discussion by taking into account the non-smooth
nature of Gaussian white noise.

It is simplest to initially consider both theories (\ref{eq:mea}) and
(\ref{eq:bodeker2a}) in $A_0=0$ gauge:
\begin {equation}
   \sigma {d\over dt} \, \A = -\D \times \B + \bnoise \,,
\label{eq:bodeker2}
\end {equation}
and
\begin {equation}
   \sigma {d\over dt} \, \A = -\D \times \B + \PT \, \bnoise \,.
\label{eq:me}
\end {equation}
For the moment, imagine a particular instantiation of the white noise
$\bnoise(\x,t)$---that is,
consider a particular member of the Gaussian ensemble
of noise functions.  Suppose that B\"odeker's equation (\ref{eq:bodeker2})
is satisfied by a gauge field $\A(\x,t)$.
Now rewrite B\"odeker's equation in the form
\begin {equation}
   \sigma \left( {d\over dt} \, \A - \sigma^{-1} \PL \bnoise \right)
        = -\D \times \B + \PT \bnoise \,.
\end {equation}
Using the explicit form (\ref{eq:PLdef}) of the longitudinal projection
operator, this can be written as
\begin {equation}
   \sigma \left( {d\over dt} \, \A - \D \tilde A_0 \right)
        = -\D \times \B + \PT \bnoise \,,
\label{eq:rewrite}
\end {equation}
where $\tilde A_0$ is simply a (suggestive) name for
\begin {equation}
   \tilde A_0 \equiv \sigma^{-1} D^{-2} \D \cdot \bnoise \,.
\label {eq:newA0}
\end {equation}

For a particular instantiation of the noise (and any initial condition),
the solution to (\ref {eq:rewrite}) may be interpreted in two different ways.
On the one hand, $\A$ is, by construction, an $A_0=0$ gauge solution
to B\"odeker's equation (\ref{eq:bodeker2a}).
On the other hand, if one says that $\tilde A_0$ is actually the
time component of the gauge field,
then the left-hand-side of (\ref {eq:rewrite}) is just $-\sigma \E$.
Therefore, $A_\mu = (\tilde A_0, \A)$
is a solution to the projected equation (\ref {eq:mea})
in the particular gauge where $A_0 = \sigma^{-1} D^{-2} \D \cdot\bnoise$.

But, given a solution $(\tilde A_0,\A)$ to (\ref{eq:mea}) with $A_0\ne0$,
one may always gauge transform back to $A_0 = 0$ gauge.
The result will be a gauge field $\barA$ which obeys
\begin {equation}
   \sigma {d\over dt} \, \barA = -\barD \times \barB + \PT \, \barbnoise .
\end {equation}
This is just the $A_0=0$ transverse equation (\ref{eq:me}),
except that the noise has been gauge transformed by the transformation
which takes $(\tilde A_0,\A)$ into $A_0 = 0$ gauge:
\begin {equation}
   \barbnoise^a = U^{ab} \bnoise^b ,
\end {equation}
with
\begin {equation}
   U = {\cal T} \exp\left[
          \int_0^t \tilde A_0 \> dt \right]
     = {\cal T} \exp\left[
          \sigma^{-1} \int_0^t {1\over D^2} \, \D\cdot\bnoise \> dt \right],
\label{eq:U}
\end {equation}
where ${\cal T}$ signifies time ordering, with the latest times on the
right.

The distinction between $\bnoise$ and its gauge transform $\barbnoise$ will
not matter, and our two theories (\ref{eq:bodeker2}) and (\ref{eq:me}) will
be equivalent (subject to earlier caveats), provided the distribution
$\barbnoise^a = U^{ab}\bnoise^b$ is Gaussian white noise,
just like the distribution of the original $\bnoise$.
If our transformation $U$ was not a
function of the noise, this would be trivial because then
we would have
\begin {equation}
   \left\langle \barbnoise^a \barbnoise^b \right\rangle
   = U^{ac} U^{bd} \, \left\langle \bnoise^c \bnoise^d \right\rangle .
\end {equation}
Since the $\bnoise$ correlator is proportional to $\delta^{cd}$, and since
$U^{ac} U^{bd} \delta^{cd} = \delta^{ab}$, it would follow that
\begin {equation}
   \left\langle \barbnoise^a \barbnoise^b \right\rangle
   = \left\langle \bnoise^a \bnoise^b \right\rangle .
\label{eq:disteq}
\end {equation}
Even though our transformation $U$ depends on the Gaussian white noise
$\bnoise$, this result naively remains true.
Consider, for instance, the equal time correlation
\begin {equation}
   \left\langle \barbnoise^a(t) \barbnoise^b(t) \right\rangle
   = \left\langle U^{ac}(t) U^{bd}(t) \, \bnoise^c(t) \bnoise^d(t)
     \right\rangle .
\end {equation}
Because the noise correlation is local in time,
while $U$ (formally) depends only on the noise {\it prior} to $t$,
this can be factorized:%
\footnote
    {%
    In fact,
    the dependence (or lack thereof) of $U(t)$ on the noise at exactly time $t$
    is ill-defined and depends on the details of time discretization,
    as discussed in the next sub-section.
    }
\begin {equation}
   \left\langle \barbnoise^a(t) \barbnoise^b(t) \right\rangle
   = \left\langle U^{ac}(t) U^{bd}(t) \right\rangle
   \left\langle \bnoise^c(t) \bnoise^d(t) \right\rangle .
\end {equation}
The $\bnoise$ correlation is again proportional to $\delta^{cd}$, which again
contracts the $U$'s and eliminates them, so that we arrive at
(\ref{eq:disteq}) as desired.  A similar argument shows that unequal time
correlations of $\barbnoise$ vanish, as they should.


\subsection{So what's wrong?}
\label{sec:wrong}

\subsubsection{A toy model}

To see what goes wrong with the previous equivalency argument, and to
understand what it has to do with the short-time nature of white noise,
it is instructive first to consider a system much simpler than
non-Abelian gauge theory.  Forget about field theory and instead imagine
stochastic dynamics of a classical particle moving in two dimensions in a
rotationally-symmetric potential $V(r)$:
\begin {equation}
   {d\over dt} \, \r = - \grad V + \bnoise ,
\label{eq:toy}
\end {equation}
\begin {equation}
   \left\langle \noise_i(t) \, \noise_j(t') \right\rangle =
   2 T \, \delta_{ij} \, \delta(t{-}t') .
\end {equation}
For convenience, we have normalized the analog of $\sigma$ to 1.
Imagine also we care only about the radial dynamics
of this system and not at all about the angular dynamics.

Comparing to the gauge theory problem, $\r$ above is analogous to $\A$
in $A_0 = 0$ gauge, the radial dynamics to the transverse gauge dynamics,
and the angular dynamics to the longitudinal dynamics.  Circles about
the origin are analogous to gauge orbits of 3-dimensional gauge
configurations under 3-dimensional gauge transformations
(since, in the gauge theory, infinitesimal displacements in the
longitudinal direction are of the form $\Delta\A(\x) = \D \Lambda(\x)$, which
is the form of an infinitesimal 3-dimensional gauge transformation).
Eq.\ (\ref{eq:toy}) is analogous to B\"odeker's effective theory
(\ref{eq:bodeker2}), and the analog of the transverse-projected theory
(\ref{eq:me}) is then
\begin {equation}
   {d\over dt} \, \r = - \grad V + \Pr \, \bnoise ,
\label{eq:toyr}
\end {equation}
where the radial projection operator $\Pr$ is
\begin {equation}
   \Pr^{ij} = \hat r^i \hat r^j = \delta^{ij} - \hat\theta^i \hat\theta^j.
\end {equation}

Just as in section \ref{sec:naive}, we can make a sloppy and not
quite correct argument that the unprojected equation (\ref{eq:toy}) and
the projected equation (\ref{eq:toyr}) are equivalent.  A transformation
from a solution $\r$ of the unprojected equation to a solution $\barr$
of the projected equation appears to be
\begin {equation}
   \barr = U \r, \qquad\qquad\qquad \barbnoise = U \bnoise ,
\end {equation}
or equivalently
\begin {equation}
   \r = U^{-1} \barr, \qquad\qquad\qquad \bnoise = U^{-1} \barbnoise ,
\label {eq:toytransform}
\end {equation}
where, if $\r$ and $\bnoise$ are represented by complex numbers
$r_x + i r_y$ and $\noise_x + i \noise_y$, $U$
can be written in a form quite analogous to (\ref{eq:U}):%
\begin {equation}
   U = \exp\left( - i \int_0^t \hat\theta \cdot \bnoise \> dt \right) .
\label{eq:toyU}
\end {equation}
$U$ simply rotates away the accumulated motion in the angular direction,
so that the projected motion, at every instant in time,
becomes purely radial.  Naively plugging
(\ref{eq:toytransform}) into the unprojected equation (\ref{eq:toy}),
and implicitly but incorrectly assuming that $\bnoise$ is a smooth
function of time, yields the naive projected equation (\ref{eq:toyr})
for $\barr$.

One can immediately see that the two equations (\ref{eq:toy}) and
(\ref{eq:toyr}) cannot, however, actually describe the same radial
dynamics.  In the unprojected case (\ref{eq:toy}), there is zero
probability that the system would ever pass {\it exactly} through
the origin $r=0$.  The projected case (\ref{eq:toyr}), however,
just describes one-dimensional motion, parameterized
by $r$, along a line of constant $\theta$.  That is, we could fix
$\theta$ and just replace (\ref{eq:toyr}) by the one degree of freedom
equation
\begin {equation}
   {d\over dt} \, r = - {dV\over dr} + \noise .
\label{eq:toy1}
\end {equation}
(There does not appear to be an analog of
this simplification in the gauge theory;
see Appendix \ref{app:notheta}.)
As long as there are no infinite potential barriers, this one-dimensional
system will
eventually fluctuate through any value of $r$, including $r=0$.

To understand the discrepancy, we need to properly understand the small
time behavior of white noise Langevin equations such as (\ref{eq:toy})
and (\ref{eq:toyr}).  The standard way of defining what such equations
actually mean is to discretize time and only at the end of the day take the
continuous time limit.


\subsubsection {Time discretization ambiguities}

\begin {figure}
\vbox
   {%
   \begin {center}
      \leavevmode
      
      \epsfbox {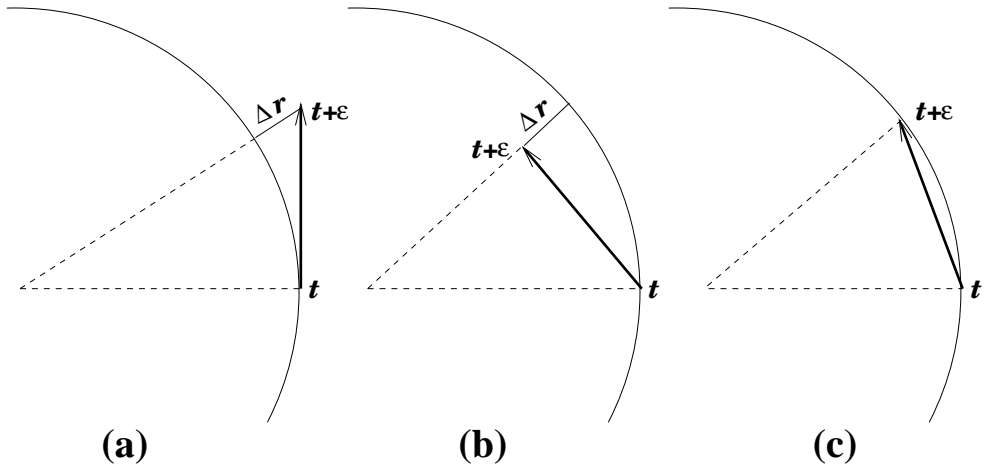}
   \end {center}
   \caption
       {%
          Radial motion generated by (\ref{eq:discreteth})
          depending on whether $\hat\theta$ is
          defined by the (a) beginning, (b) end, or (c) middle of a
          discrete time jump $t \to t+\eps$.
       \label{fig:ambiguity}
       }%
   }%
\end {figure}

Before proceeding, we have to dispose of an instructive
red herring concerning the
time discretization of our various stochastic equations.
It is not always true that continuum-time stochastic equations like the
ones we have been writing down have an unambiguous meaning.  To understand
the possible ambiguities, imagine that instead of being interested in only
the radial dynamics of our toy model, we were instead interested in only the
angular dynamics, and so had proposed a projected equation of the form
\begin {equation}
   {d\over dt} \r = P_\theta \, \bnoise ,
\end {equation}
\begin {equation}
   P_\theta^{ij} =
      \hat \theta^i \hat \theta^j = \delta^{ij} - \hat r^i \hat r^j.
\end {equation}
This continuum equation {\it appears}\/ to describe motion for which
the radius $r$ remains constant.
Now imagine discretizing time with small time steps of size $\eps$, so
that
\begin {equation}
   \eps^{-1} \Delta \r = P_\theta \, \bnoise ,
\label{eq:discreteth}
\end {equation}
\begin {equation}
   \langle \noise_i(t) \, \noise_j(t{+}n\eps) \rangle
      = 2 T \eps^{-1} \, \delta_{ij} \delta_{n0} .
\label {eq:discretenoise}
\end {equation}
There is an ambiguity in the schematic way we have written the discretized
equation (\ref{eq:discreteth}): we have not made it clear whether the
direction $\hat\theta$ implicit in $P_\theta$ is supposed to be
evaluated at the starting point of the tiny time interval, the end point,
or somewhere in between.  In the first case, the value of $r$ will drift out a
little bit, as in fig.\ \ref{fig:ambiguity}a.
In the second case, it will drift in a little
bit, as in fig.\ \ref{fig:ambiguity}b.
If we pick a symmetric convention, where we evaluate
$\hat\theta$ at the midpoint, then $r$ will remain constant, as in
fig.\ \ref{fig:ambiguity}c.

In non-stochastic equations, such discretization choices
become irrelevant in the continuum limit $\eps \to 0$ (though they may
have significance for the practicality of numerical calculations).
For stochastic
equations, however, the $\eps\to 0$ limit is more subtle because,
by (\ref{eq:discretenoise}), the
amplitude of the white noise $\bnoise$ is order
\begin {equation}
    \noise \sim \sqrt{T \over \eps}
\end {equation}
and diverges as $\eps \to 0$.
The drift $\Delta r$ in figs.~\ref{fig:ambiguity}a and
b is therefore of order
\begin {equation}
   \pm \Delta r \sim \sqrt{r^2 + (\eps \noise)^2} - r
                \sim {(\eps \noise)^2 \over r} \sim {\eps T \over r}
\end {equation}
for a time interval $\eps$.  That means that the drift per unit time,
$\Delta r/\eps$, is finite as $\eps \to 0$, and so the continuum limit
really depends on one's discretization conventions.

In the unprojected equation (\ref{eq:toy}), there is no such
discretization ambiguity.  And in our actual toy model equation
(\ref{eq:toyr}) with {\it radial} projection, there is no such
ambiguity because motion in the $\hat\r$ direction, unlike in the $\hat\theta$
direction, is straight---$\hat\r$ does not change between one end of the
interval and the other.  The situation is slightly more complicated for
the transverse-projected equation (\ref{eq:me}) for gauge theory,
however.  There, motion of $\A$ in the transverse direction {\it does}
change the transverse projection operator $\PT$.  However, we demonstrate
in section \ref{sec:vanish} that this change
turns out to be high enough order in $\delta\A$ that discretization
ambiguities do not arise.

The upshot is that the continuum stochastic equations (\ref{eq:toyr})
and (\ref{eq:me}) for the radial-projected toy model and the
transverse-projected gauge theory are not ambiguous.  However, we shall
next see that the very same discretization issues affect the transformations
we used to argue that they were equivalent with their unprojected
counterparts.


\subsubsection {Centrifugal drift}
\label{sec:toy_correction}

The way we proposed obtaining the projected equation (\ref{eq:toyr})
from the unprojected one (\ref{eq:toy}) was by rotating away the
accumulated $\theta$ motion.  Imagine a single time step of the
discretized unprojected equation.  Then%
\footnote{
  Whether it's $V(r(t))$ or $V(r(t{+}\eps))$, or the average of the two,
  does not matter in the continuum time limit $\eps \to 0$.
}
\begin {equation}
    \r(t+\eps) = \r(t) - \eps \grad V(r(t)) + \eps \bnoise(t) .
\end {equation}
The motion of the radial coordinate $r$ is then
\begin {eqnarray}
    r(t+\eps) &=& \left|\r(t) - \eps \grad V(r(t)) + \eps \bnoise(t)\right|
\nonumber
\\ &=&
     \left.\left[ r - \eps V'(r) + \eps \, \hat\r \cdot \bnoise
		+ {\eps^2\over 2r} |\bnoise|^2 \right]\right|_t
		+ O(\eps^{3/2}) .
\end {eqnarray}
Given that, as $\eps \to 0$, a large number of successive tiny time steps
will occur before the system appreciably changes position, the positive
$|\noise|^2$ term can be replaced by its statistical average
(\ref{eq:discretenoise}):
\begin {equation}
    r(t+\eps)
         \approx \left[ r - \eps V'(r) + \eps \, \hat\r \cdot \bnoise
                    + {\eps T \over r} \right](t) .
\label{eq:boyoboy}
\end {equation}
The distribution of $\hat\r\cdot\bnoise$ does not care about the direction
of $\hat\r$, and (\ref{eq:boyoboy}) can be rewritten as
\begin {equation}
   r(t+\eps) \approx r(t) - \eps V'_\eff(r(t)) + \eps \noise_r(t) ,
\end {equation}
where $\noise_r$ is uncorrelated white noise
and
\begin {equation}
   V_\eff(r) = V(r) - T \ln r .
\label{eq:Veff}
\end {equation}
The continuum projected equations that are truly equivalent to the
unprojected one are therefore (\ref{eq:toyr}) or (\ref{eq:toy1})
with $V$ replaced by $V_\eff$.  The addition of the $\ln r$ term in
(\ref{eq:Veff}) now provides a ``centrifugal potential'' which prevents
the one-dimensional system (\ref{eq:toy1}) from passing through $r=0$.


\subsubsection {Equilibrium distributions}
\label{sec:toyP}

The exact form of the ``centrifugal'' correction was really determined from
the very start.  As we shall briefly review in
section \ref{sec:gaugeP}, the equilibrium distribution in $\r$
generated by the unprojected equation (\ref{eq:toy1}) is proportional
to $\exp(-V/T)$.  That means that the probability distribution
for the radial variable $r$ must be proportional to
\begin {equation}
   2\pi r \, \exp(-V/T) \propto \exp(-V_\eff/T) ,
\label{eq:dist1}
\end {equation}
since the $2\pi r$ is the volume of the symmetry orbit.
But $\exp(-V_\eff/T)$
is precisely the equilibrium distribution generated by the projected
equation (\ref{eq:toy1}) if $V$ is replaced by $V_\eff$.

In the gauge theory case, there will be no analog to the one-dimensional
radial equation (\ref{eq:toy1}), and so it is worthwhile to understand
how the equilibrium distribution could have been deduced directly from
the two-dimensional projected equation (\ref{eq:toyr}).
This requires deriving the Fokker-Planck equation
that is associated with a given Langevin equation and which describes
the time evolution of probability distributions $\Prob(\r)$.
(This will be discussed explicitly in the gauge theory case below.)
For the naive projected equation (\ref{eq:toyr}), one finds that
\begin {equation}
   \Prob(\r) \propto {\exp(-V/T) \over 2 \pi r} 
\label{eq:distr0}
\end {equation}
is a (two-dimensional) equilibrium solution.
But if we correct the naive equation by replacing $V \to V_\eff$,
then we indeed recover the unprojected result
\begin {equation}
   \Prob(\r) \propto {\exp(-V_\eff/T) \over 2\pi r}  = \exp(-V/T)
\label{eq:distr}
\end {equation}
as the equilibrium distribution in $\r$.%
\footnote
    {%
    Because the projected equation conserves $\theta$, 
    there is a family of two-dimensional equilibrium solutions
    to the projected toy model equation:
    namely, the rotationally invariant distribution (\ref{eq:distr0})
    can be multiplied by an arbitrary angular distribution $f(\theta)$.
    This non-uniqueness is, of course, irrelevant if one is only
    interested in rotationally invariant observables.
    The appearance of an arbitrary function of $\theta$ in the general
    equilibrium distribution is a reflection of the non-ergodicity
    of the projected two-dimensional evolution equation.
    As discussed in Appendix \ref {app:notheta},
    for the transverse-projected gauge theory there does not appear
    to be any analog of a conserved gauge-orbit coordinate $\theta$
    and, so far as we know, the transverse-projected gauge theory
    remains ergodic.
    }
[Do not confuse the two-dimensional distribution (\ref{eq:distr}) for $\r$
with the one-dimensional radial distribution (\ref{eq:dist1}) for $r$.
Both describe the same equilibrium ensemble.]


\subsection {Gauge Theory}

\subsubsection {Time discretization ambiguities}

The transverse-projected version (\ref{eq:me}) of the soft effective theory
is a Langevin equation of the form
\begin {mathletters}%
\label{eq:b1}%
\begin {equation}
   {d\over dt} \, q_i = - \partial_i V(\q) + e_{ia}(\q) \, \noise_a ,
\end {equation}%
\begin {equation}
   \langle\noise_a(t) \, \noise_b(t')\rangle
      = 2 {\cal T} \, \delta_{ab} \, \delta(t{-}t') ,
\end {equation}%
\end{mathletters}%
where, for the moment, we are using notation natural for a system with
a finite number of degrees of freedom.  In the field theory case, the
dynamical variables $\q$ are the values of the gauge fields at different
points in space and $\partial_i$ becomes a functional derivative
$\delta/\delta A$.
The functions $e_{ia}(\q)$ characterize the coupling of the noise
to the dynamical variables $\q$;
for the gauge theory this is the transverse projection
operator $\PT$ (which depends on the gauge field~$\A$).

To define exactly what is meant by this equation, imagine
discretizing time into very small time steps of size $\epsilon$.%
\footnote{
  The following discussion roughly follows the presentation in
  sections 4.7 and 4.8 of ref.\ \cite{ZinnJustin},
  although our normalizations are slightly different.
}
Stochastic equations of the form (\ref{eq:b1}) are generically
ambiguous if the coupling $e_{ia}(\q)$ to the noise has non-trivial
dependence of $\q$, because of the ambiguity, discussed earlier,
of when to evaluate $\q$.
In the discretized equation,
\begin {equation}
   {q_i(t+\eps) - q_i(t) \over \eps} =
   \left[-\partial_i V + e_{ia} \, \noise_a \right]_{t+\alpha\eps} ,
\end {equation}
\begin {equation}
   \langle\noise_a(t) \, \noise_b(t')\rangle
    = 2 {\cal T} \eps^{-1} \, \delta_{ab} \, \delta_{t t'} \,,
\end {equation}
this ambiguity appears as dependence on a parameter $\alpha$
which controls the time at which the right-hand side is evaluated.
For example,
$\alpha=0$ corresponds to a forward time derivative and is known
as the It\^o convention, $\alpha=\half$ to the symmetric
derivative, known as Stratonovich convention, and $\alpha=1$ to a
backward time derivative.
The precise meaning of evaluation at time $t+\alpha\eps$ is to
expand in $\alpha\eps$.
Keeping in mind that
the amplitude of the noise $\noise$ is $\eps^{-1/2}$, and using the equation
of motion itself, the terms
in the expansion which are non-negligible when $\eps\to0$ are
\begin {equation}
   {q_i(t+\eps) - q_i(t) \over \eps} =
   \Bigl[ -\partial_i V + e_{ia} \, \noise_a
      + \alpha \, \epsilon \, (\partial_j e_{ia}) e_{jb} \, \noise_a \noise_b
   \Bigr]_t .
\end {equation}
The product
$\noise_a \noise_b$ may be replaced by its expectation, giving the final
discretized equation
\begin {equation}
   {q_i(t+\eps) - q_i(t) \over \eps} =
   \Bigl[ -\partial_i V
               + 2\alpha {\cal T} \, (\partial_j e_{ia}) e_{ja}
          + e_{ia} \, \noise_a \Bigr]_t .
\label{eq:b2}
\end {equation}
The term proportional to $\alpha$ is a convention-dependent ``drift''
force.
The naive continuous-time formulation (\ref{eq:b1})
does not, in general, uniquely specify the dynamics.


\subsubsection {Vanishing ambiguity}
\label{sec:vanish}

We shall now show that the ambiguity vanishes for the transverse-only
noise equation of (\ref{eq:me}).  This implies that there is no real
issue of convention dependence for this application.

We work in continuous space (rather than working on a spatial lattice,
which would be more relevant to numerical simulations but also more
complicated).
The degrees of freedom in the gauge theory case are labeled by
coordinates $X = (\x,i,a)$, where $i$ is a vector index and $a$ an adjoint
color index.
It will be convenient to introduce combined labels for several different
choices of coordinates:
\begin {equation}
     X = (\x,i,a), \qquad\qquad Y = (\y,j,b), \qquad\qquad Z = (\z,k,c) .
\end {equation}
The noise coupling $e_{ia}$ introduced above becomes the
(matrix elements of the) transverse projection operator
\begin {equation}
   P_{XY} =
        \delta^{ij}\delta^{ab}\,\delta(\x{-}\y) - (D^i D^{-2} D^j)^{ab}_{\x\y}
   .
\label{eq:bP}
\end {equation}
This operator is symmetric in $X$ and $Y$, and the drift force discussed
earlier is proportional to
\begin {equation}
        P_{XZ} \, {\delta\over\delta A_X} \, P_{ZY} .
\end {equation}
When taking the variation of $P_{ZY}$, the variation must act on the left-most
covariant derivative in (\ref{eq:bP}), since otherwise that derivative
will annihilate against the $P_{XZ}$ factor.  One thus obtains
\begin {eqnarray}
     P_{XZ} \, {\delta\over\delta A_X} \, P_{ZY}
     &\propto& \int_\z P_{XZ} \,
             \delta^{ik} \, T^a_{ce} \, \delta(\x-\z) \,
             \left(D^{-2} D^j\right)^{eb}_{\z\y}
\nonumber\\
     &\propto& d \, T^a_{ae} \left(D^{-2} D^j\right)^{eb}_{\x\y}
             - (D^i D^{-2} D^i)^{ac}_{\x\x} \, T^a_{ce}
             \left(D^{-2} D^j\right)^{eb}_{\x\y}
\end{eqnarray}
in $d$ spatial dimensions,
where no integration over $\x$ is implied.
The first term vanishes because the adjoint generators $T^a_{bc}$ are
anti-symmetric in $(abc)$ and so $T^a_{ae}=0$.  The second
term vanishes because $(D^i D^{-2} D^i)^{ac}_{\x\x}$ is symmetric in
$(ac)$ and so vanishes when contracted with the anti-symmetric
$T^a_{ce}$.  So the convention-dependent drift force is exactly zero.%
\footnote{
   The drift force is also proportional to ${\cal T}$ which, in the field
   theory case, is $T \delta({\bf 0})$.  If we were only interested in
   perturbative physics, we could have chosen to work in dimensional
   regularization, which sets $\delta({\bf 0})$ to zero.
}%
\thinspace\footnote{
\label{fn:general}
  In the general case, a sufficient condition for the ambiguity to vanish can be
  expressed as follows.  Suppose the potential $V(\q)$ of (\ref{eq:b1})
  is invariant under some symmetry transformations
  that have the infinitesimal form
  $\q \to \q + \lambda \btheta^\alpha(\q)$, where
  $\alpha$ indexes the independent symmetry generators
  (and $\lambda$ is infinitesimal).
  Define a metric $g^{\alpha\beta} = \btheta^\alpha \cdot \btheta^\beta$
  on the space of symmetry
  generators.  As in general relativity, let $g_{\alpha\beta}$
  denote the (matrix)
  inverse of the metric.
  Now suppose that the noise coupling equals the
  projection operator
  $e_{ij} = \delta_{ij} - \theta_i^\alpha g_{\alpha\beta} \, \theta^\beta_j$.
  One can then show that the ambiguity
  $(\partial_j e_{ia}) e_{ja}$ vanishes if
  $\partial_i \theta^\alpha_j$ is anti-symmetric under interchange
  of $i$ and $j$.
  This anti-symmetry condition is indeed satisfied in both the
  radial-projected toy model
  and transverse-projected gauge theory.
}


\subsubsection {Centrifugal drift}

We now return to the transformation,
between B\"odeker's effective
theory and the transverse-projected theory,
in order to derive the gauge theory analog of the centrifugal correction
discussed for the toy model in section~\ref{sec:toy_correction}.
The time-discretized version of B\"odeker's effective theory is
\begin {equation}
    \A(t{+}\eps) = \A(t)
      - {\eps \over \sigma} \,
      \left({\delta\V\over\delta\A} (t) - \bnoise(t) \right) \,,
\label{eq:Adisc}
\end {equation}
where
\begin {equation}
   \V = \half \, g^{-2} \int_\x B^a_i B^a_i =
   \fourth \, g^{-2} \int_\x F^a_{ij} F^a_{ij}
\end {equation}
is the potential energy associated with the magnetic field,
the noise correlation is
\begin {equation}
    \left\langle \noise_i^a(t,\x) \, \noise^b_j(t{+}n\eps,\x') \right\rangle
      = 2 g^2 T \sigma \eps^{-1} \delta^{ab} \delta_{ij} \delta_{n0} \,
            \delta(\x{-}\x') \,,
\label{eq:Adiscnoise}
\end {equation}
and we have chosen to use a forward time difference.
We now want to apply (a discrete version of) the gauge transformation
$U$ that was introduced in the time-continuum case, (\ref{eq:U}),
to eliminate longitudinal motion of $\A$.
For simplicity of presentation, we shall focus on one single,
time step from $t$ to $t {+} \epsilon$ and discuss how
to transform $\A(t{+}\eps)$ relative to $\A(t)$ in order to eliminate
the longitudinal motion introduced during that step.
Consider a gauge transformation $U$ which equals the identity
at time $t$, but is a non-trivial infinitesimal transformation
at time $t {+} \eps$,
\begin {equation}
    U(t,\x) = 1, \qquad U(t{+}\eps,\x) = \exp\alpha(\x) \,.
\end {equation}
For the moment, we leave the generator of the transformation, $\alpha$,
arbitrary.
The gauge transformed field is
\begin {equation}
   \barA \equiv U (\grad + A) U^{-1} \,.
\end {equation}
Expanding in powers of the generator $\alpha$ at
time $t{+}\eps$, this gives
\begin {equation}
    \barA (t{+}\eps)
    = \A(t{+}\eps) - \D\alpha + \half[\D\alpha,\alpha]
    + O(\alpha^3) ,
\end {equation}
where $\D\alpha = \grad\alpha + [\A(t{+}\eps),\alpha]$.
Using the equation of motion (\ref{eq:Adisc})
to rewrite $\A(t{+}\eps)$ in terms of $\A(t)$ gives
\begin {equation}
   \barA(t{+}\eps)
   = \A(t)
    - {\eps\over\sigma}\left({\delta\V\over\delta\A} - \bnoise\right)
    - \D\alpha
    + {\eps\over\sigma}\left[{\delta\V\over\delta\A} - \bnoise,\alpha\right]
    + \half [\D\alpha,\alpha]
    + O(\alpha^3,\sqrt\eps \alpha^2) ,
\label {eq:barEOM}
\end {equation}
where now all the covariant derivatives involve the gauge field at time $t$.
Choosing the infinitesimal generator to equal
\begin {equation}
   \alpha = {\eps\over\sigma} \, D^{-2} \, \D\cdot\bnoise \,,
\end {equation}
as implied by our previous discussion [{\em c.f.}\ (\ref{eq:U})],
will cause the $\D\alpha$ term to cancel the longitudinal projection
of the noise $\bnoise$.
Since the noise $\bnoise$ is of order $\eps^{-1/2}$, this means that
$\alpha$ is of order $\sqrt \eps$.
We need to keep all terms in (\ref {eq:barEOM}) which are of order $\eps$.
The term $(\eps/\sigma) \, [\delta \V,\alpha]$ is $O(\eps^{3/2})$
and may be neglected.
Consequently,
\begin {eqnarray}
   \barA(t{+}\eps) = \A(t)
    &+& {\eps\over\sigma}\left(-{\delta\V\over\delta\A} + \PT \bnoise\right)
    - {\eps^2\over\sigma^2}
   \left[\bnoise - \half \D D^{-2} \D\cdot\bnoise, D^{-2} \D\cdot\bnoise\right]
    + O(\eps^{3/2}).
\end {eqnarray}
Once again, we can replace the terms quadratic in noise by
their statistical averages, as given by (\ref{eq:Adiscnoise}).
After some manipulation, one finds that this yields
\begin {eqnarray}
   \barA^a(t{+}\eps,\x) &=&
    \A^a(t,\x)
    + {\eps\over\sigma}\left(-{\delta\V\over\delta\A^a(\x)}
    + \PT \, \bnoise^a(t,\x)\right)
    + {T \eps\over\sigma} \, f^{abc} \, (\D \, D^{-2})_{\x\x}^{cb}
    + O(\eps^3)
\nonumber\\ &=&
    \A^a(t,\x)
    + {\eps\over\sigma}\left(-{\delta\V_\eff\over\delta\A^a(\x)}
    + \PT \, \bnoise^a(t,\x)\right)
    + O(\eps^3) ,
\end {eqnarray}
where
the effective potential $\V_\eff$ is
\begin {equation}
   \V_\eff = \V - \half T \, \Tr \ln (-D^2)
           = \V - T \ln \sqrt{\det (-D^2)} \,.
\end {equation}
As shown in Appendix~\ref{app:volume},
$\sqrt{\det(-D^2)}$ is the volume of the gauge orbit
containing a given spatial gauge field.
Consequently, this logarithmic correction to the potential is
completely analogous to the ``centrifugal'' potential appearing
in the rotationally invariant toy model.
The upshot is that the projected equation which is actually equivalent
to B\"odeker's effective theory differs from the naive projected
equation (\ref{eq:me}) by the replacement of $\V$ by $\V_\eff$:
\begin {equation}
    \sigma {d\over dt} \, \A =
    -\D \times \B + \half T {\delta \over \delta\A} \, \Tr \ln(-D^2)
    + \PT \, \bnoise \,.
\label{eq:mecorrect}
\end {equation}


\subsubsection {The equilibrium distribution}
\label {sec:gaugeP}

It's interesting to examine what happens
if one converts a Langevin equation of the generic form (\ref{eq:b2})
into a Fokker-Planck
equation for the evolution of the probability distribution $\Prob(\q,t)$.
One finds (see for example \cite{ZinnJustin}) that
\begin {equation}
   {\partial\over\partial t} \, \Prob
   = \partial_i
   \Bigl[ \,
       {\cal T} \partial_j \left( e_{ia} e_{j a} \Prob \right)
        + \left\{ \partial_i V - 2\alpha {\cal T} \,
                (\partial_j e_{ia}) e_{ja} \right\} \Prob
    \,\Bigr] .
\end {equation}
If $e_{ia}(\q)$ were just $\delta_{ia}$, as in B\"odeker's
equation (\ref{eq:bodeker2}), then the equilibrium distribution
(the solution to $d\Prob/dt = 0$)
would simply be $\Prob = \exp(-V/{\cal T})$ up to an overall
normalization constant.

For the naive radial-projected toy model (\ref{eq:toyr}) with
$e_{ia} = \hat r_i \hat r_a$,
\begin {equation}
   \partial_j (e_{ia} e_{ja}) = {\hat r_i \over r}
       = \hat r_i {d\over dr} \ln (2 \pi r) ,
\end {equation}
and $(\partial_j e_{ia}) e_{ja} = 0$.  This leads to the
equilibrium distributions (\ref{eq:distr0}) quoted in
section \ref{sec:toyP}.

In the gauge theory case, we have seen that the convention-dependent drift
force vanishes for the transverse-only noise equation (\ref{eq:me}),
but there still
remains $e_{ia}$ dependence in the Fokker-Planck equation.
Plugging in the transverse projection operator for $e$, one finds
\begin {equation}
   \partial_j(e_{ia} e_{ja})
   \to {\delta\over\delta A_Y} \, P_{XY}
   = \half P_{XY} {\delta\over\delta A_Y} \,\tr \ln(-D^2) ,
\end {equation}
and, solving for the equilibrium distribution,
\begin {equation}
   \Prob = {\exp(-\V/T) \over \sqrt{\det(-D^2)}} 
\end {equation}
up to an overall normalization constant.%
\footnote{
   In the more general notation of footnote \ref{fn:general},
   the assumption that $\partial_i \theta^\alpha_j$ is anti-symmetric
   in $i$ and $j$ leads to
   $\partial_j(e_{ia} e_{ja}) = \half e_{ij} \partial_j \ln \sqrt{g}$
   and
   $\Prob = \exp(-V/T) / \sqrt{g}$,
   where $g$ is the determinant of the inverse metric
   $g_{\alpha\beta}$.
}
As mentioned above, $\sqrt{\det(-D^2)}$ just
represents the gauge orbit volume, and the
above distribution is analogous to the toy model result
(\ref{eq:distr0}).  As with the toy model, however, if we examine
the transverse theory that is truly equivalent to the unprojected
theory, then we should replace $\V \to \V_\eff$ and we recover the
correct equilibrium distribution
\begin {equation}
   \Prob = {\exp(-\V_\eff/T) \over \sqrt{\det(-D^2)}} 
         = \exp(-\V/T) \,.
\end {equation}


\section* {ACKNOWLEDGMENTS}

We thank Dietrich B\"odeker and Guy Moore
for useful conversations.
We are especially indebted to Deitrich B\"odeker for
conversations, concerning an earlier version of this manuscript,
which inspired us to clarify our understanding of the
$\omega \ll k \ll \gamma_\g$ limit of the longitudinal sector.
This work was supported, in part, by the U.S. Department
of Energy under Grant Nos.~DE-FG03-96ER40956 and DF-FC02-94ER40818.


\appendix

\section {The volume of gauge orbits}
\label{app:volume}

The natural metric on the space of gauge field vector potentials
is
\begin {equation}
    ds^2 = \int_\x \> \tr \left[ d\A^\dagger \cdot d\A \right] .
\label {eq:metric}
\end {equation}
This is the unique metric (up to an overall multiplicative constant)
which is invariant under both gauge transformations and spacetime symmetries.
The gauge orbit passing through
a particular gauge configuration $\A$
consists of all gauge transforms of $\A$.
Within a neighborhood of $\A$, configurations on the gauge orbit
may be parameterized as
\begin {equation}
    \A^\Lambda \equiv e^{-\Lambda} \, \D \, e^\Lambda \,,
\end {equation}
where $\Lambda$ is an arbitrary generator of the gauge group ({\em i.e.},
$\Lambda(\x) \equiv \Lambda^a(\x) \, T^a$ is a Lie-algebra valued function
of $\x$),
and $\D = \grad + \A$ is the covariant derivative with gauge field $\A$.
Since $\A^\Lambda - \A \sim \D\Lambda$,
the induced metric on the gauge orbit, evaluated at $\A$, is just
\begin {equation}
    \left. ds^2 \right|_{\rm orbit}
    =
    \int_\x \>
    \tr \left[(\D \delta\Lambda)^\dagger \cdot (\D \delta\Lambda)\right]
    =
    \int_\x \> \tr \left[\delta\Lambda^\dagger (-D^2)\, \delta\Lambda\right] .
\end {equation}
Consequently the induced volume element on the orbit,
evaluated at $\A$, is
\begin {equation}
    dv = \sqrt{ \det (-D^2)} \> d\Lambda \,,
\end {equation}
where $d\Lambda \equiv \prod_{\x,a} d\Lambda^a(\x)$ denotes the flat measure
on the gauge algebra.
But the gauge-invariant Haar measure on the gauge group is just
$d\Lambda$ when evaluated at the identity.
And, because the functional determinant $\det (-D^2)$ is gauge invariant,
it is constant over the gauge orbit.
So, globally, the volume element $dv$ equals
$\sqrt {\det (-D^2)}$ times the Haar measure on the gauge group.
Hence,
\begin {equation}
    {\hbox {orbit volume} \over \hbox {gauge group volume}}
    =
    \left[ \det (-D^2) \right]^{1/2} \,,
\end {equation}
and so $\sqrt {\det(-D^2)}$ is the gauge orbit volume
up to an overall $\A$-independent normalization factor.

\section {No gauge theory analog to toy model $\theta$}
\label{app:notheta}

Return, for a moment, to the toy model described in Sec.\ \ref{sec:wrong}.
The projected equation (\ref{eq:toyr}) has two obvious properties.
First, the particle always moves in a direction perpendicular to the gauge
orbits $r=\text{const}$.  Second, moving according to this equation, the
particle cannot reach any point in the configuration space, but instead
remains confined to a slice in a configuration space where $\theta$ is a
constant. In particular, starting from a point $(r,\theta)$, one cannot
reach any point that is gauge equivalent to it, except the original point. 
In other words, if one fixes the gauge $\theta=\theta_0$, with $\theta_0$
some constant, then this gauge-fixing condition remains satisfied throughout
the random walk. 

Now consider the gauge theory.  Eq.\ (\ref{eq:me}) describes a
motion in the space of field configurations that is analogous
to that described by the projected equation (\ref{eq:toyr}) in the toy model.
In terms of the natural metric (\ref {eq:metric}),
on the space of gauge configurations,
one can easily check that the motion is always along directions perpendicular
to deformations generated by gauge transformations
(This is equivalent to satisfying the condition $\D\cdot\dot{\A}=0$.)
The question we want to ask is whether the second
property of our toy model still holds, {\em i.e.},
is the motion confined to a slice in configuration space?
It might be surprising that the answer to this question is negative,
and one can travel throughout the whole configuration space
even when restricted to trajectories whose tangents, at every point,
are perpendicular to the intersecting gauge orbit.
This negative answer is perhaps 
less surprising if one notices that there is no obvious gauge-fixing condition
similar to $\theta=\theta_0$ that is conserved during the
transverse-projected random walk (\ref{eq:me}).  Thus, in the gauge
theory, there is no equivalence of the parameter $\theta$. 

This can be seen most directly by the explicit construction of a
trajectory that remains perpendicular to gauge transformations
at all times,
but nevertheless connects two distinct points on the same gauge orbit.
The trajectory we
are going to construct starts at $\A=0$ and remains small all the time, so
we can use perturbation theory.  Let us denote the small parameter by
$\eps$.  Consider first the following trajectory,
\begin{equation}
  A_i(t,\x) = \left\{ \begin{array}{ll} 
    tC_i(\x)                         \,,\quad& 0 < t < \eps; \\
    \eps C_i(\x) + (t{-}\eps) D_i(\x)  \,,\quad& \eps < t < 2\eps; \\
    (3\eps{-}t) C_i(\x) + \eps D_i(\x) \,,\quad&  2\eps < t < 3\eps; \\
    (4\eps{-}t) D_i(\x) \,,\quad& 3\eps < t < 4\eps \,.
    \end{array} \right.
  \label{app:Aleading}
\end{equation}
Provided that $C_i$ and $D_i$ are transverse,
$\partial_iC_i=\partial_iD_i=0$, it is trivial to check that
$\D\cdot\dot{\A}=0$ to leading order in $\eps$.  This means the
trajectory is everywhere perpendicular (to within $O(\epsilon^2)$)
to the gauge orbits it passes through.

This trajectory may seem uninteresting, since it is a
closed loop that starts at $\A=0$ and ends at the same point.
The interest arises when we go to next-to-leading order in $\eps$.
At next-to-leading order, the trajectory
(\ref{app:Aleading}) does not satisfy the condition $\D\cdot\dot{\A}=0$. 
For example, when $\epsilon<t<2\epsilon$, $\D\cdot\dot{\A}=\eps[C_i,D_i]$.
To correct for this deviation, we need to modify $A_i$ in the following
way:
\begin{equation}
  A_i(t,\x)  = \eps C_i(\x) + (t-\eps) D_i(\x)+(t-\eps)\alpha_i(\x), 
  \qquad \eps < t < 2\eps\,, \\
  \label{app:Acorrected}
\end{equation}
where $\alpha_i= O(\eps)$, so the term involving $\alpha_i$ is of higher
order than the other terms.  Eq.\ (\ref{app:Acorrected}) satisfies the
condition $\D\cdot\dot{\A}=0$ if one places the following constraint on
$\alpha_i$:
\begin {equation}
  \partial_i\dot{\alpha}_i + \eps [C_i, D_i] = 0 \,.
\end {equation}
This condition can be satisfied by choosing
\begin {equation}
  \dot{\alpha}_i = - \eps \, \partial_i\grad^{-2} [C_j, D_j] \,.
\end {equation}
In this manner, one can modify the whole trajectory (\ref{app:Aleading}) 
so that the condition $\D\cdot\dot{\A}=0$ is satisfied through
next-to-leading order.
The result is
\begin{equation}
  A_i(t,\x) = \left\{ \begin{array}{ll}
    tC_i                                       \,,\quad&  0 < t < \eps ;\\
    \eps C_i + (t{-}\eps) D_i - 
    \eps(t{-}\eps) \partial_i\grad^{-2}[C_j,D_j] \,,\quad& \eps < t < 2\eps ;\\
    (3\eps{-}t) C_i + \eps D_i - 
    (\eps^2 + \eps(t{-}2\eps)) \partial_i\grad^{-2}[C_j,D_j]
                                               \,,\quad& 2\eps < t < 3\eps ;\\
    (4\eps{-}t) D_i -
     2\eps^2 \partial_i\grad^{-2}[C_j,D_j]     \,,\quad& 3\eps < t < 4\eps \,.
    \end{array} \right.
\end{equation}
One sees that the trajectory now starts at $A_i=0$ at $t=0$ and runs to
$A_i=-2\eps^2 \partial_i\grad^{-2}[C_j,D_j]$ at $t=4\eps$.
[Including still higher-order corrections would only change this
by $O(\epsilon^3)$.]
At the end point, $\A$ is a pure gauge, and,
for a general choice of $C_i$ and $D_i$, nonzero.
Therefore, this trajectory presents a simple example of how,
following a transverse projected trajectory,
one can go from one field configuration to a field configuration that is gauge
equivalent to it.
 From this result one may show
(provided the gauge group is semi-simple) that
{\em any} field configuration in the
vicinity of $\A=0$ is accessible to the transverse-projected random walk
(\ref{eq:me}).
Hence, there can be no analog of
the toy model
``slice parameter'' $\theta$
for transverse-projected dynamics in non-Abelian gauge theories.



\end {document}